\documentclass{skaox2006}
\usepackage{graphicx}
\begin{document}
   \title{The ALMA Observation Support Tool}

   \author{I.~Heywood\inst{1}, A.~Avison$^{2}$
          \and
           C.~J.~Williams$^{3}$
          }

   \institute{$^{1}$ Subdepartment of Astrophysics, University of Oxford, Denys-Wilkinson Building, Keble Road, Oxford, OX1 3RH, UK \\
	$^{2}$ Jodrell Bank Centre for Astrophysics, The University of Manchester, Oxford Road, Manchester, M13 9PL, UK \\
	$^{3}$ Oxford e-Research Centre, 7 Keble Road, Oxford, OX1 3QG, UK\\
	}

   \abstract{
   The ALMA Observation Support Tool (OST) is an ALMA simulator which is 
interacted with solely via a standard web browser. It is aimed at users who 
may or may not be experts in interferometry, or those that do not wish to 
familarise themselves with the simulation components of a data reduction 
package. It has been designed to offer full imaging simulation capability 
for an arbitrary ALMA observation while maintaining the accessibility of 
other online tools such as the ALMA Sensitivity Calculator. Simulation jobs 
are defined by selecting and entering options on a standard web form. The 
user can specify the standard parameters that would need to be considered 
for an ALMA observation (e.g. pointing direction, frequency set up, 
duration), and there is also the option to upload arbitrary sky models in 
FITS format. Once submitted, jobs are sequentially processed by a remote 
server running a CASA-based back-end system. The user is notified by email 
when the job is complete, and directed to a standard web page which 
contains the results of the simulation and a range of downloadable data 
products. The system is currently hosted by the UK ALMA Regional Centre, 
and can be accessed by directing a web browser to {\tt 
http://almaost.jb.man.ac.uk}.
}
   
   \authorrunning{Heywood et al.}
   \titlerunning{ALMA OST}
   
   \maketitle
   
%
%

\section{Overview}

The Atacama Large Millimetre/submillimetre Array (ALMA)\footnote{{\tt http://www.almatelescope.org}} is an interferometer consisting of 66 dishes
currently under construction on the Chajnantor plateau of northern Chile. Operating between $\sim$90 GHz and $\sim$1 THz, 
it will be the most sensitive instrument in the world at these frequencies when completed in 2012. Observations with a 16-element ALMA will begin in 2011. 

The Observation Support Tool (OST) provides a new method for simulating ALMA images. It is designed to be easily accessible to users who may 
not be interferometry experts, and is interacted with solely via a standard web browser. 
No additional software needs to be installed by the user, and in contrast to other web-based tools such as the ALMA Sensitivity 
Calculator\footnote{{\tt http://almascience.eso.org/document-and-tools}} no client-side processing takes place. Instead, 
simulation jobs are defined via a standard web form and submitted to a remote server. The server is running a custom-built script which processes
the submitted jobs sequentially, making use
of the CASA\footnote{{\tt http://casa.nrao.edu}} toolkit to perform a full simulation via the generation and imaging of a visibility set. When the simulation is complete the user receives an
email containing a URL which points to a web page containing the results of the simulation and several downloadable image products.

The front end makes use of an open source Javascript form checking library called LiveValidation\footnote{{\tt http://www.livevalidation.com}}
which checks the simulation parameters in real time. Basic errors can thus be trapped before jobs are submitted to the server. A second level of 
server-side error checking is employed to test for more complex issues (e.g. sources remaining below the horizon at all times) and these error
checking routines also replicate the LiveValidation checks for users who choose to disable Javascript in their browsers.

This article provides a brief overview of the functionality of the OST by describing the web front-end and the results page. The OST is currently
hosted by the UK ALMA Regional Centre at the University of Manchester Jodrell Bank Centre for Astrophysics, and can be accessed by directing a web
browser to {\tt http://almaost.jb.man.ac.uk}.

\section{The web interface}

\begin{figure*}
\vspace{428pt}
\includegraphics{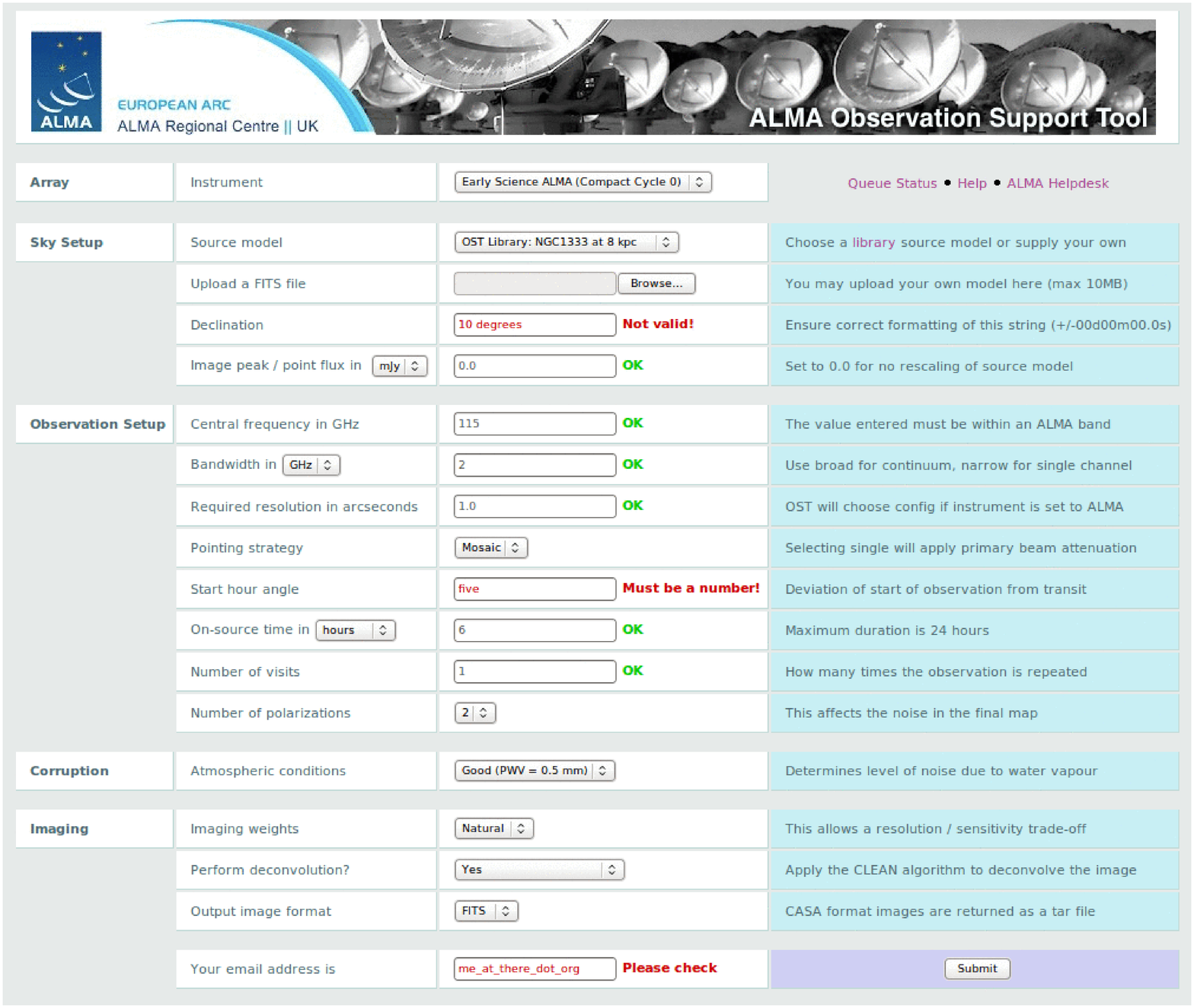}
\caption{The web interface of the OST, showing the LiveValidation library in action.
        \label{fig:front_end}
}
\end{figure*}

The interface is a standard HTML web form with various components of the simulation defined either by text boxes or drop-down menus as shown in Figure \ref{fig:front_end}. 
It is divided
into five main sections, as defined by the leftmost column. The column on the right-hand side provides brief notes as to the purpose and usage of each 
item, however (evolving) documentation
is also available from a hyperlink on the web page. The red and green markers show the LiveValidation library in action, with erroneous parameters highlighted in red.
A brief summary of each section on the web interface follows.

\subsection{Array}

The instrument is defined here using a single drop-down menu. The available options are two Cycle-0 configurations (compact and extended), the
12-element Atacama Compact Array (ACA) and the full 50-element ALMA. Although the full ALMA array will have numerous potential configurations
there is only one option for it in this menu. The reasons for this are explained in Section \ref{sec:obs_setup}.

\subsection{Sky Setup}
\label{sec:sky_setup}

This section determines the nature of the `ideal' sky that will be used as the input model for the simulation. The OST contains
a library of example sky models which may be used, and flexibility as a simulation tool
is provided in this section as the user is able to upload an arbitrary sky model in FITS format. The only header parameter in the FITS file that the user must ensure
is accurate is the spatial pixel scale. Everything else is ignored by the OST and is instead overwritten by the values defined elsewhere in the web form, however
if the brightness unit keyword is Jy/pixel or Jy/beam then this is also taken into account. 

Right Ascension is conspicuous by its absence next to the Declination field, however this was foregone in favour of an hour angle parameter (see Section 
\ref{sec:obs_setup}).

The final option in this section can be used to re-scale the pixel values of the sky model by setting the brightness of the peak pixel
and scaling all the other pixel values in the image array in relation to this. 

\subsection{Observation Setup}
\label{sec:obs_setup}

The spectral and temporal properties of the observation are defined here. The OST at present only does single-channel simulations, however the
increased sensitivity offered by large-bandwidth pseudo-continuum
observations can be simulated by simply entering a large bandwidth, together with the central frequency. For spectral line simulations the user unfortunately
at present has to submit each channel as a unique simulation job, however selecting a few representative channels across the model spectral line cube is suitable
for a crude line detection test. Frequency cubes can be uploaded, however the OST will only make use of the central channel. The user will be notified on the results
page if this action is taken.

The required resolution parameter is only taken into account by the simulation algorithm if the user selects `ALMA' in the Array section. The required resolution is
used in conjunction with the central frequency to select a suitable ALMA configuration from the 28 `out' layouts which are bundled with CASA. If the resolution
demand falls outside the range of what is offered by the most compact or extended arrays then one of the extreme configurations is selected and the user is notified.

A major observing mode for ALMA will be mosaicking. In mosaic mode, the OST takes the sky area demands of the model into account, and calculates a standard hexagonal
mosaic pattern\footnote{Pointings in any given row are offset by half the primary beam width, and adjacent rows are offset horizontally from one-another by
one-quarter of the primary beam width, and vertically by the primary beam width multipled by $sin$(60 degrees)}. The OST will calculate the number of pointings required
to cover the requested area, but it also makes the assumption that the specified on source time is per pointing, and not divided amongst them. The other option
in this section is to simulate a single pointing, whereby a pixel-wise attenuation is applied the sky model pre-simulation, crudely simulating the primary beam response
of the array. The attenuation function is a normalized Gaussian, the centre of which is placed at the central pixel of the sky model. Its full-width at half-maximum is 
an angular value, equal in radians to
\begin{equation}
\Theta_{PB}~=~1.22\frac{c}{\nu D}
\end{equation}
where $c$ is the speed of light, in metres per second, $\nu$ is the central observing frequency in Hertz and $D$ is the dish diameter in metres.

As mentioned previously, the Right Ascension parameter for the model sky is not present, and instead the user must specify a starting hour-angle value. Thus the simulation is
defined in terms of when the source is observed in relation to its transit. 

The pointing duration is subsequently specified, and there is also the option to specify a number of visits. This allows the user to simulate cases whereby a large amount
of observing time is required but the hour angle ranges are stringent.

Finally the number of polarizations can be selected. The OST does not yet perform simulations in full polarization, and this parameter merely affects the noise in the
final image. As with spectral lines however, individual polarization maps can be uploaded. This prospect also partially motivated the allowance of negative
brightness values in the re-scaling option, although if a negative value is entered here then the user is warned on the results page in case it was unintentional.

\subsection{Corruption}
\label{sec:corruption}

Artifacts in an interferometric image derived from a real observation originate due to a variety of effects, including calibration errors, atmospheric effects
and the thermal conditions of the receivers. Even in the case of a perfect observation, the latter is something that cannot be mitigated. The thermal noise
determines the absolute noise `floor' in an interferometric map below which sources cannot be detected. 

The RMS of the noise perturbation to the visibility, that is the single complex number which is the per-polarization, per-channel correlation product of a pair of antennas, is given in units of Janskys by:
\begin{equation}
\label{eq:rms}
\sigma = \frac{2k_{B}T_{sys}}{\eta A \sqrt{\Delta \nu \Delta t}}
\end{equation}
where $k_{B}$ is the Boltzmann constant, $T_{sys} = T_{rec} + T_{sky} + T_{cmb}$ is the system temperature in K, $\eta$ is the combined product of a series of 
efficiency terms
$A$ is the effective area of a single antenna in m$^{2}$, $\Delta \nu$ is the channel bandwidth in Hz and $\Delta t$ is the integration
time per visibility in seconds.

The receiver temperatures have a unique value per ALMA band and $T_{cmb}$ is set to 2.73 K. 
The sky temperature
is derived from a model of the atmospheric transparency at the ALMA site via:
\begin{equation}
\label{eq:tsky}
T_{sky} = T_{atmos} \left(1 - \gamma\right)
\end{equation}
where $T_{atmos}$ is the atmospheric temperature (assumed to be 260 K) and $\gamma$ is the transmission fraction.

\begin{figure*}
\vspace{250pt}
\includegraphics{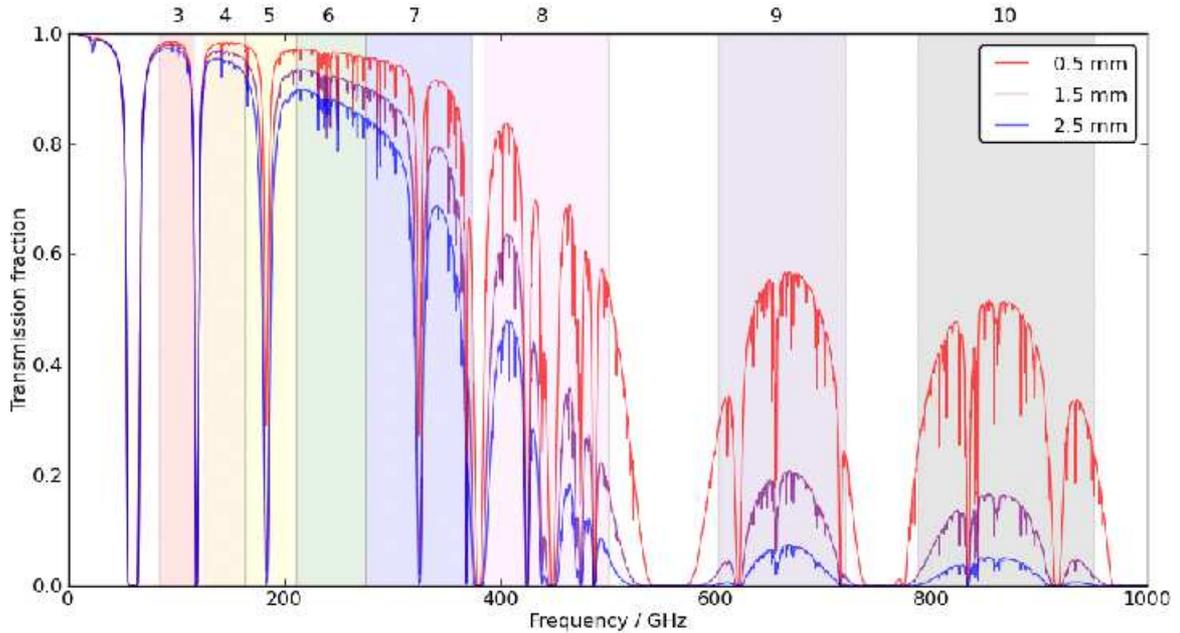}
\caption{Atmospheric transmission fraction at the ALMA site as a function of frequency for three column densities of precipitable
 water vapour. The vertical bars and the numbers above the plot denote ALMA bands 3--10.
        \label{fig:trans}
}
\end{figure*}

The single menu option in the `Corruption' section relates to three levels of precipitable water vapour (PWV): 0.5, 1.5 and 2.5 mm. The atmospheric
transmission fraction as a function of frequency for these three levels is shown in Figure \ref{fig:trans}. The values in this plot have
been derived from the transmission calculator on the Atacama Pathfinder Experiment (APEX) web site\footnote{{\tt http://www.apex-telescope.org/sites/chajnantor/atmosphere}}.

 Interpolative functions are fitted to these plots so that the transparency can be derived for arbitrary frequencies and the corresponding sky
 temperature can be calculated via Equation \ref{eq:tsky} for the selected
 level of PWV. To account for the variation of $T_{sky}$ within large bandwidths, the values are
 calculated at ten points across the band and an average is taken. This value is then added to $T_{rec}$ to form $T_{sys}$ for use in
 Equation \ref{eq:rms}.

\subsection{Imaging}

The parameters in this section affect how the simulated visibilites are Fourier transformed into a sky image which then undergoes optional
deconvolution. The imaging process for an interferometric data set does not necessarily follow a single unique path, thus complete automation of this process is a challenge.
How the gridded visibilites are weighted can affect the final map, with weighting schemes offering a trade-off between resolution and sensitivity. 
The OST offers three weighting options: natural weighting, pure uniform weighting and an intermediate Briggs (1995) weighting scheme.

When imaging a genuine observation, deconvolution is often carried out interactively, allowing the user to adaptively define regions to be targeted 
by the CLEAN algorithm and make an educated call as to when deconvolution should be terminated. If the OST user wishes to deconvolve
the simulated map then the OST will set a termination threshold according to the theoretical noise in the map. When the peak of the residual image
reaches this value then deconvolution ceases. If the deconvolution process fails to converge or otherwise fails to reach the threshold then it will be terminated by means
of a clean component limit which is presently set to 10,000.

The final output image format option in this section simply determines whether the OST returns the downloadable data products in FITS format or the CASA image format.

\begin{figure}
\centering
\includegraphics[width=\columnwidth]{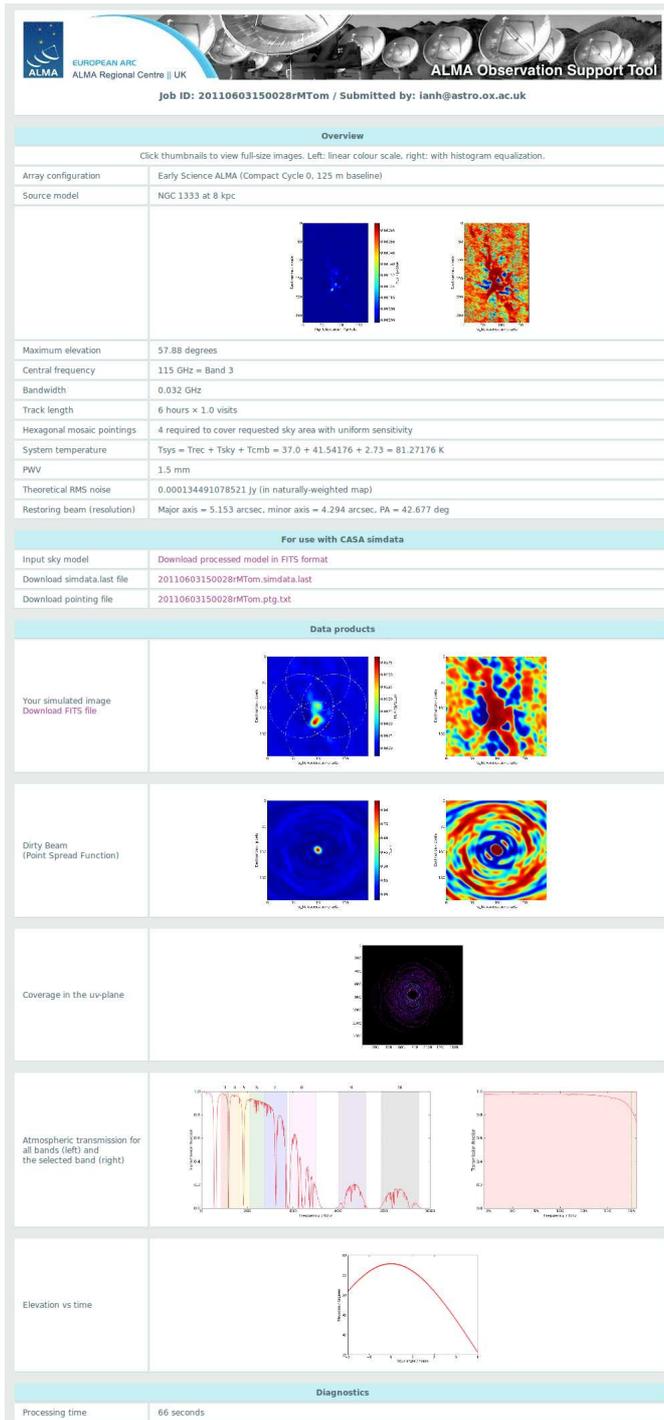}
\caption{An example of the results page.}
\label{fig:results}
\end{figure}

\section{The results page}

Once the simulation has been processed the OST will send the user an email containing a link to a web page similar to that shown in Figure \ref{fig:results}.
The content of this page is influenced by user feedback and is subject to change at the time of writing. The results page is also divided into four sections.
Sometimes an extra section will appear at the head of the page containing warning messages. These generally occur when the OST has encountered a non-fatal problem
with the simulation and has had to take liberties with the simulation parameters. One example of this would be the requested bandwidth causing the frequency range
to spill over a band edge, in which case the OST will truncate the frequency coverage and notify the user.

Images are rendered in PNG format and are presented with both a linear pixel intensity scale (on the left) and with histogram 
equalization\footnote{Histogram equalization is an image processing technique which adjusts the pixel intensity histogram of an image such that it has a flat distribution. This
is particularly useful for enhancing low level structure in images where the dynamic range is governed by the presence of a few very bright pixels.}
(on the right).

\begin{figure}
\centering
\includegraphics[width= 0.9 \columnwidth]{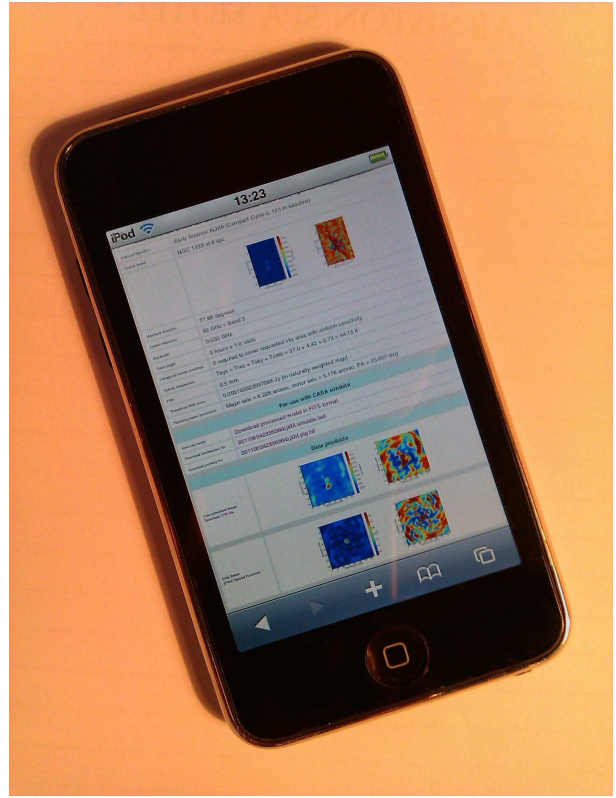}
\caption{Using the OST on an iPod Touch. Ease of access is one of the key features of the simulator.}
\label{fig:ipod}
\end{figure}

\subsection{Overview} 

A few useful parameters about the simulation are presented here, some of which are repeats of what is entered into the webform and some of which are derived (e.g. maximum
elevation, resolution of final map, number of pointings). If the simulation used a non-point-source sky model then this is rendered and displayed here.

\subsection{CASA simdata}

This section provides three downloads which are designed to allow the user to easily transfer their OST simulation into the CASA simdata task. The
processed sky model is offered as a download and the simdata.last file
will set up simdata with the parameters used for the OST simulation by means of the CASA tget command. The pointing file contains a list of the directions for each pointing in the mosaic. 
Note however that as mentioned earlier the OST uses an hour angle parameter instead of Right Ascension, thus the RA of every OST simulation is forced to zero
hours.

If the `single' option is chosen for the pointing strategy then in addition to the processed sky model the sky model with primary beam attenuation applied 
is also returned in the chosen image format. Such features may be useful to users who wish to ignore the simulation components
and simply use the OST as an online FITS file re-processing service.

\subsection{Data products}

The data products section contains solely graphical output. The most useful image here is probably the rendering of the final simulated map. For mosaics of less
than 30 pointings the pointing directions are overlaid onto this image. The link to download the simulated map as either a FITS or CASA image is also here.
The dirty beam (or point-spread function) of the observation is also presented. The $uv$ coverage of the simulation is displayed here. This is generated by Fourier 
transforming the PSF rather than opening the visibility set and extracting the $uvw$ coordinates of each measurement. The advantage of this approach is that it is much faster,
and the colour scale in the $uv$ plot gives some indication as to the density of samples in a particular region of the $uv$ plane.

The frequency set up is distilled into the fourth row of plots, showing the frequency range of the simulation in the context of ALMA bands 3--10, and the band
in which it lies. The red atmospheric transmission curve also reflects the PWV level that was selected. All three possible transmission curves are shown in Figure \ref{fig:trans}
of this document.

Finally this section also presents a plot of elevation against time. Scans are flagged when an antenna drops below an elevation angle of 10 degrees.
Such scans are displayed in a lighter tone on the elevation plot so the user can easily see the fraction of their observation which is affected. A message
will appear on the results window notifying the user of any elevation issues, and noise values calculated from the on-source time are also scaled accordingly.

\subsection{Diagnostics}

This section contains only a single result, which is the number of seconds that elapse between the simulation job being selected from the queue and the 
job being completed. Turn-around times are very favourable, although the back-end is being modified to exploit the multiple CPU cores of the server,
allowing jobs in the queue to be processed in parallel rather than the current sequential implementation.

\section{Summary}

The Observation Support Tool is a flexible imaging simulator for ALMA which is highly accessible, the software requirements it places upon its users consisting of
nothing more than a standard web browser (e.g. Figure \ref{fig:ipod}). The server-side software which processes the simulations make use of the CASA toolkit, and although the OST
is not a web interface to the CASA simdata task, both systems are based on this toolkit and further good agreement is ensured via careful matching of assumed parameters.

The OST itself also serves as an example of the potential of using remote computing applicatons for radio astronomical data processing. The new generation of radio
interferometers (e.g. EVLA, e-MERLIN, LOFAR and eventually MeerKAT, ASKAP and the Square Kilometre Array) require high-end computers to perform data 
reduction, and the use of remotely-accessed HPC facilities is likely to become increasingly important, an approach which has already been adopted for the
processing of LOFAR data which takes place on a central cluster facility.

The system is at present in a state of constant refinement as its performance is checked and user feedback is received. The system is completely open
for general use and has to date processed over 1,500 simulations.

\begin{acknowledgements}
The authors would like to thank the users of the OST, especially the testers who volunteered to try out an early version of the system and whose
feedback and bug detections proved invaluable, in particular Eduardo Ibar, Robert Laing, Fran\c{c}ois Levrier, Tom Muxlow, 
and Anita Richards. We are also very grateful to Remy Indebetouw and the CASA development teams at NRAO and ESO.
We thank STFC for financial support via the ALMA ARClet grant.

\end{acknowledgements}

\end{document}